\documentclass[pra,twocolumn,floatfix,showpacs,superscriptaddress,nofootinbib]{revtex4}

\usepackage{hyperref}
\usepackage{hypernat}
\usepackage{times,amsmath,amsfonts,amssymb,latexsym}
\usepackage{graphicx}
\usepackage{epsfig}

\begin{document}

\title{Channel Capacities of an Exactly Solvable Spin-Star System}
\author{Nigum Arshed}
\affiliation{Department of Physics, Quaid-i-Azam University Islamabad 45320, Pakistan}
\author{A. H. Toor}
\affiliation{Department of Physics, Quaid-i-Azam University Islamabad 45320, Pakistan}
\author{Daniel A. Lidar}
\affiliation{Departments of Physics, Chemistry, and Electrical Engineering, Center for
Quantum Information Science \& Technology, University of Southern
California, Los Angeles 90089, USA}

\begin{abstract}
We calculate the entanglement-assisted and unassisted channel capacities of
an exactly solvable spin star system, which models\ the quantum dephasing
channel. The capacities for this non-Markovian model exhibit a strong
dependence on the coupling strengths of the bath spins with the system, the
bath temperature, and the number of bath spins. For equal couplings and bath
frequencies, the channel becomes periodically noiseless.
\end{abstract}


\maketitle

\section{Introduction}

One of the fundamental tasks of quantum information theory is to determine
the information transmission capacities of quantum channels \cite{Bennett:98review,Nielsen:book}. The maximum amount of information that can
be reliably transmitted over a channel, per channel use is known as its 
\textit{capacity }\cite{Cover:book}. Classical channels can be uniquely
characterized by their capacity \cite{Shannon:48}. The situation in the
quantum realm is significantly more involved, with various capacities
required to characterize a quantum channel \cite{Bennett:04}.

Studies of quantum channel capacities can be broadly divided into those
considering memoryless quantum channels, for which the output at a given time
depends only upon the corresponding input and not upon any previous inputs 
\cite{Hausladen:96,Schumacher:97,Lloyd:97a,Bennett:97,Adami:97,Barnum:98,Bennett:99,Bennett:02,Shor:02,Holevo:02,Liang:02,Daffer:03a,Shor:04,Kretschmann:04,Devetak:05,Giovannetti:05a,Wolf:07}, and quantum memory channels, where successive uses of the channel modify
its properties and description \cite{Macchiavello:02,Yeo:03,Macchiavello:04,Bowen:04,Arshed:06,Arrigo:07,Plenio:07,Bayat:08}. Another important distinction is between channels generated by Markovian
{\it vs}. non-Markovian environments or baths. Markovian channels describe
memoryless baths, while for non-Markovian channels bath memory plays a role 
\cite{Breuer:book,Breuer:09}. Many quantum optical \cite{Carmichael:93} and nuclear
magnetic resonance systems \cite{Slichter:book} are accurately described by
Markovian channels, but the Markovian limit is always an approximation \cite{AlickiLidarZanardi:05}. Non-Markovian effects are especially important in
condensed matter systems, such as coupled electron or nuclear spins \cite{Weiss:book}. The master equations describing the dynamics of non-Markovian
systems are often (though not always \cite{ShabaniLidar:05}) complicated
integro-differential equations which are rarely exactly solvable \cite{Breuer:book}.
A channel can be memoryless yet non-Markovian. This situation arises
when successive uses do not modify the channel, but a proper description of
each use of the channel requires a non-Markovian treatment accounting for
bath memory effects.
In this work we investigate how non-Markovian effects modify channel
capacities by studying an exactly solvable model of a non-Markovian
memoryless channel: the Ising spin-star system \cite{Krovi:07}.

One reason to consider spin systems is that they are good candidates for the
physical realization of quantum computation and communication,
in part due to their relatively long relaxation and decoherence times \cite{Loss:98,Kane:98,Vrijen:00,Petta:05,Morton:08}. Spin chains have attracted
much recent interest as quantum communication channels \cite{Bose:07}.
Capacities of a spin chain with ferromagnetic Heisenberg interactions were
calculated by studying the qubit amplitude damping channel \cite{Giovannetti:05a}, and its successive use without resetting (quantum memory
channel) was investigated for quantum and classical communication \cite{Bayat:08}.

Different flavors of the spin-star system, with both diagonal and
non-diagonal coupling, have been used to study topics such as entanglement
distribution \cite{Hutton:04}, the dynamics of entanglement of two central
spins \cite{Yuan:07}, and analytically solvable models of decoherence
\cite{Krovi:07,Breuer:04a,Hamdouni:06}. However, spin-star systems are so far
unexplored for quantum transmission of information,
and this is our goal in the present paper. The system qubit in our communication model is represented by a spin
located at the center of the star. It interacts with all non-central spins,
comprising the environment, via Ising couplings. This provides a
non-Markovian quantum dephasing channel whose dynamics can be solved exactly 
\cite{Krovi:07}. We allow arbitrary couplings between the system and
environment spins and unlike the spin chain channels studied in Refs. \cite%
{Giovannetti:05a,Bayat:08}, obtain analytical expressions for the capacities
of this model. We do not consider the quantum memory channel
setting of successive channel uses, wherein a new spin is repeatedly introduced into the same
channel \cite{Arrigo:07}. Rather, we consider the parallel use setting
of a memoryless channel \cite{Giovannetti:05a},
where $n$ messages (classical or quantum) are simultaneously transmitted over $n$\ identical spin-star
systems. Thus, in our treatment, non-Markovian memory effects are
entirely associated with the non-Markovian dynamics
of each spin-star system.

The organization of the paper is as follows. In Sec.~\ref{sec:cap} we give a
brief review of quantum channels and their capacities. In Sec.~\ref{sec:deph}
we describe the model of a quantum dephasing channel obtained by coupling a
system spin via Ising interactions to a spin bath, and review its exact
solution in the Kraus representation.
In Sec.~\ref{sec:comm} we present our communication model, calculate its
capacities and study some limiting cases. Finally, in Sec.~\ref{sec:conc} we
discuss the results and present our conclusions. Appendix~\ref{app}
contains a technical calculation.

\section{Quantum Channel Capacities}

\label{sec:cap}

Formally, a quantum channel $\mathcal{E}$\ is a completely positive and
trace preserving map (CPTP) of a quantum system from an initial system state 
$\rho _{S}$ to the final state $\mathcal{E}( \rho _{S}) $ \cite{Nielsen:book,Breuer:book}. Quantum channels arise by joint unitary
evolution $U$ of the system and its environment or bath, followed by a
partial trace ${\rm Tr}_{B}$ over the bath, if and only if system and bath start
from a purely classically correlated initial state \cite{ShabaniLidar:08},
such as a product state:%
\begin{equation}
\rho _{S}\mapsto \mathcal{E}( \rho _{S}) =\text{Tr}_{B}[
U( \rho _{S}\otimes \rho _{B}) U^{\dagger }] .
\label{Quantum Channel}
\end{equation}%
Here $\rho _{B}$ is the initial state of the bath. The conjugate $\widetilde{\mathcal{E}}$ of a quantum channel $\mathcal{E}$ is defined as \cite{King:07},
\begin{equation}
\widetilde{\mathcal{E}}( \rho _{B}) =\text{Tr}_{S}[ U(
\rho _{S}\otimes \rho _{B}) U^{\dagger }] .
\label{Conjugate Channel}
\end{equation}%
A quantum channel\ is called degradable if it can be degraded to its
conjugate, that is, there exists a CPTP map $\mathcal{T}$ such that $\widetilde{\mathcal{E}}=\mathcal{T}\circ \mathcal{E}$ \cite{DevetakShor:05}.
We shall make use of degradable channels later on in this work.

Unlike classical channels at least four capacities are associated with
quantum channels depending on the type of information transmitted (classical
or quantum), protocols allowed, and auxiliary resources used \cite{Bennett:04}. We are interested in the classical capacity $C$, quantum
capacity $Q$, and entanglement-assisted capacities $C_{E}$, $Q_{E}=C_E/2$ and $C_{E}^{\lim }$ of a quantum dephasing channel.

Let $S( \rho ) =-{\rm Tr}[ \rho \log _{2}\rho ] $ denote
the von Neumann entropy. The maximum amount of classical information
reliably transmitted over a quantum channel is given by its {\em classical capacity} $C$ \cite{Hausladen:96,Schumacher:97}, 
\begin{eqnarray}
C &=&\lim_{n\rightarrow \infty }\frac{C_{n}}{n},\quad C_{n}=\max_{p_{i},\rho
_{S,i}\in \mathcal{H}_{S}^{\otimes n}}\chi  \label{Classical Capacity} \\
\chi &=&S[ \mathcal{E}^{\otimes n}( \rho _{S}) ]
-\sum_{i}p_{i}S[ \mathcal{E}^{\otimes n}( \rho _{S,i})
] .
\end{eqnarray}%
It depends on the largest set of orthogonal input states distinguishable
during the transmission and not on the ability of a channel $\mathcal{E}$ to
preserve phases of different superpositions. $C$ is the Holevo information $%
\chi $\ \cite{Holevo:73} maximized over all possible input ensembles $\rho
_{S}=\sum_{i}p_{i}\rho _{S,i}$, where $\{p_{i}\}$ is a probability
distribution and $\{\rho _{S,i}\}$ a set of quantum states
(\textquotedblleft quantum alphabet\textquotedblright\ belonging to the $n$%
-fold tensor product of system Hilbert spaces $\mathcal{H}_{S}$), in the
limit $n\rightarrow \infty $ of parallel or successive channel uses. The
limit can be avoided when the Holevo information is additive over 
channel uses, in which case the optimal ensembles which achieve the maximum
in Eq.~(\ref{Classical Capacity}) are separable with respect to the $n$ uses
and $C$ coincides with $\frac{C_{n}}{n}$ for all $n$, and in particular with 
$C_{1}$. Hastings recently provided counterexamples to the additivity of the
minimum output entropy \cite{Hastings:09}, which implies by a result of Shor
that the classical capacity is not always additive \cite{Shor:04a}.

The \textit{quantum capacity} $Q$\ is the maximum amount of quantum
information transmitted by a quantum channel per channel use \cite{Lloyd:97a,Shor:02,Devetak:05}, 
\begin{eqnarray}
Q &=&\lim_{n\rightarrow \infty }\frac{Q_{n}}{n},\quad Q_{n}=\max_{\rho
_{S}\in \mathcal{H}_{S}^{\otimes n}}I_{c}  \label{Quantum Capacity} \\
I_{c} &=&S[ \mathcal{E}^{\otimes n}( \rho _{S}) ] -S%
[ ( \mathcal{E}^{\otimes n}\otimes \mathcal{I}) (|\Phi
\rangle \langle \Phi |)] .  \label{I_c}
\end{eqnarray}%
For a given number\ of channel uses $n$, it depends on the dimension of the
largest Hilbert subspace of $\mathcal{H}_{S}^{\otimes n}$ that does not
decohere during transmission. The quantum capacity $Q$\ is the coherent
information $I_{c}$ \cite{Schumacher:97},\ maximized over all possible input
states. In Eq.~(\ref{Quantum Capacity}), $|\Phi \rangle \in \mathcal{H}%
_{S}\otimes \mathcal{H}_{R}$ is a purification of $\rho _{S}$ obtained by
appending a reference Hilbert space $\mathcal{H}_{R}$ to the system Hilbert
space $\mathcal{H}_{S}$. The limit $n\rightarrow \infty $ is necessary as $%
I_{c}$ is super-additive \cite{Barnum:98},\ which makes the evaluation of $Q$
difficult. However, for \emph{degradable }channels\ the coherent information$%
\ I_{c}$ reduces to the conditional entropy,\ which is subadditive and
concave, from which it follows that for these channels $Q=Q_{1}$
(single-channel use) \cite{DevetakShor:05}. This is an important
simplification, which enables the explicit calculation of $Q$ in a variety
of interesting cases.

Entanglement is a useful resource in quantum information transmission. For
example, it can be used to enhance the performance of quantum error
correcting codes \cite{Brun:06}, to enhance quantum channel capacities by sharing
entanglement between sender and receiver prior to communication \cite{BW92},
or by encoding information into entangled states when making successive uses
of the same channel
\cite{Macchiavello:02,Yeo:03,Macchiavello:04,Bowen:04,Arshed:06,Plenio:07,Bayat:08}.
If the sender and receiver share unlimited prior entanglement, the
maximum 
amount of classical information reliably transmitted over the quantum
channel is given by its {\em entanglement-assisted classical capacity} $%
C_{E}$ \cite{Bennett:99,Bennett:02,Holevo:02}. This quantity is obtained by
maximization of the quantum mutual information for \emph{single} channel
use, which yields%
\begin{equation}
C_{E}=\max_{\rho _{S}\in \mathcal{H}_{S}} \{ S( \rho _{S}) +S%
[ \mathcal{E}( \rho _{S}) ] -S[ ( \mathcal{E}%
\otimes \mathcal{I}) (|\Phi \rangle \langle \Phi |)] \} .
\label{Entanglement-assisted classical capacity}
\end{equation}%
Here $|\Phi \rangle \in \mathcal{H}_{S}\otimes \mathcal{H}_{R}$ is the
shared entangled state, which is also a purification of the input state $%
\rho _{S}\in \mathcal{H}_{S}$. The amount of pure-state entanglement
consumed by this communication protocol is $S(\rho _{S})$ ebits
per channel use, where $\rho _{S}$ maximizes Eq.~(\ref{Entanglement-assisted
classical capacity}). In contrast to the classical and quantum capacities, $%
C_{E}$ is additive \cite{Holevo:02}. The \textit{entanglement-assisted
quantum capacity} is given by $Q_{E}=C_{E}/2$, and can be attained by
superdense coding \cite{BW92}, and quantum teleportation \cite{Bennett:93}.

Shor has given a trade-off curve showing the classical capacity as a
function of the amount of entanglement shared by the sender and receiver 
\cite{Shor:04}. The end points of this curve are given by the classical
capacity $C$ and the entanglement-assisted classical capacity $C_{E}$. If
the amount of entanglement available $P$ is\ less than $S( \rho
_{S}) $, then the classical capacity assisted by limited entanglement
is given by%
\begin{eqnarray}
C_{E}^{\lim } &=&\max_{\{ \rho _{S,i},p_{i}\}
}\sum_{i}p_{i}S( \rho _{S,i}) +S[ \mathcal{E}(
\sum_{i}p_{i}\rho _{S,i}) ] \notag \\
&&-\sum_{i}p_{i}S[ ( \mathcal{E}\otimes I) (|\Phi_i\rangle \langle \Phi_i|)] ,
\label{Classical capacity with limited entanglement}
\end{eqnarray}%
subject to $\sum_{i}p_{i}S( \rho _{S,i}) \leq P$. Here the
maximization is over the probabilistic ensemble $\{ \rho
_{S,i},p_{i}\} $ where $\rho _{S,i}\in \mathcal{H}_{S}$, $%
\sum_{i}p_{i}=1$, $p_{i}\geq 0$, and as above the shared entangled states $%
|\Phi_i\rangle $ are purifications of $\rho _{S,i}$. The
capacity $C_{E}^{\lim }$\ reduces to the classical capacity $C$ given by Eq.
(\ref{Classical Capacity}) for $P=0$, as the constraint $\sum_{i}p_{i}S%
( \rho _{S,i}) \leq P$ implies that $\rho _{S,i}$ must then all
be pure states. For sufficiently large $P$ it gives the
entanglement-assisted classical capacity $C_{E}$. The proof of
additivity of $C_{E}^{\lim }$ is an open problem.

\section{Quantum Dephasing Channel}

\label{sec:deph}

\subsection{The Model}

\begin{figure}[tph]
\centering
\includegraphics[width=3.4in]{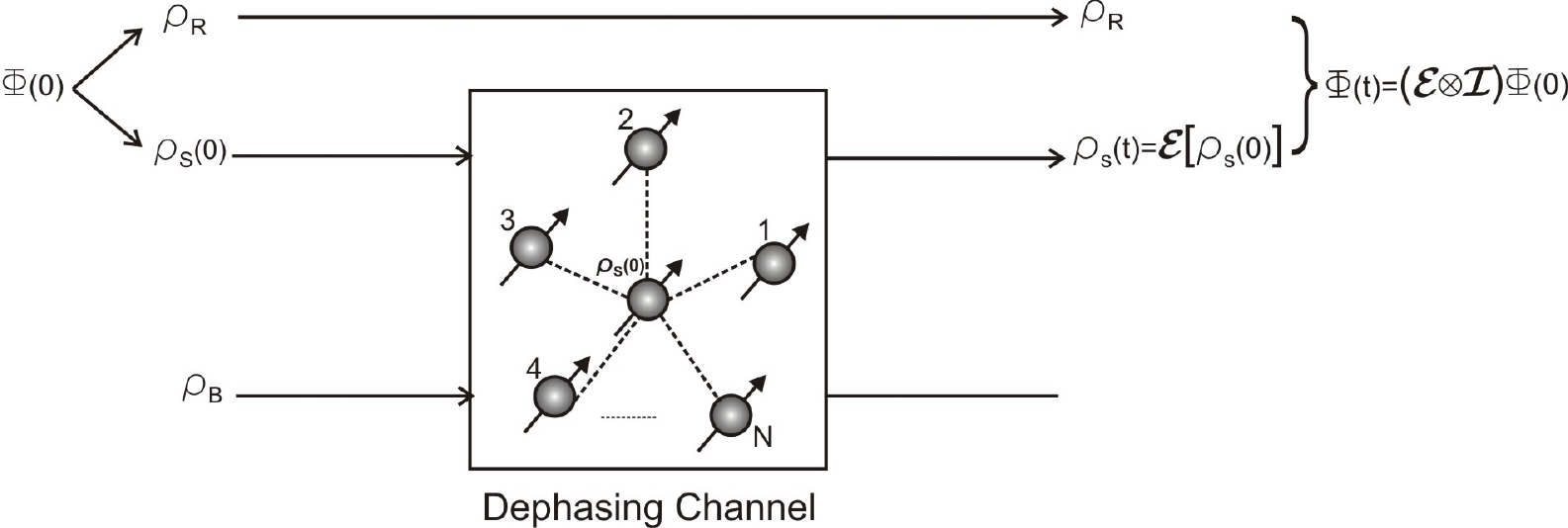}
\caption{Communication protocol: a qubit passes through an Ising spin-star
channel. This is repeated in parallel over many identical such channels.}
\label{Communication protocol}
\end{figure}

We consider the case of an exactly solvable \textit{spin star} system of $%
N+1 $ localized spin-$\frac{1}{2}$ particles as shown in
Fig.~\ref{Communication protocol}. The system input state $\rho
_{S}( 0) $ 
is carried by the central (system) spin. This spin interacts with $N$
noncentral spins comprising the bath. The bath spins do not interact with
each other directly. The interaction between the system spin and bath is
given by the Ising Hamiltonian%
\begin{equation}
H_{I}=\alpha \sigma _{z}\otimes \sum_{n=1}^{N}g_{n}\sigma _{n}^{z},
\label{Ising}
\end{equation}%
where we work in $\hbar =1$ units, $g_{n}\in [ -1,1] $ are
dimensionless real-valued coupling constants, and $\alpha >0$ is the
coupling strength having the dimension of frequency. The system and bath
Hamiltonians are given by%
\begin{eqnarray}
H_{S} &=&\frac{1}{2}\omega _{0}\sigma ^{z},  \label{System} \\
H_{B} &=&\frac{1}{2}\sum_{n=1}^{N}\Omega _{n}\sigma _{n}^{z}.  \label{Bath}
\end{eqnarray}%
The frequencies $\omega _{0}$ and $\Omega _{n}$ are restricted to the
interval $[ -1,1] $, in frequency units. Initially, the total
system is assumed to be in the product state

\begin{equation}
\rho ( 0) =\rho _{S}( 0) \otimes \rho _{B},
\label{initial state}
\end{equation}%
with the bath in the Gibbs thermal state at inverse temperature $\beta
=1/(kT)$ given by 
\begin{equation}
\rho _{B}=\exp ( -\beta H_{B}) /\text{Tr}[ \exp (
-\beta H_{B}) ] .  \label{Gibbs state}
\end{equation}%
Since $\rho _{B}$ commutes with $H_{I}$ the bath state is stationary
throughout the dynamics: $\rho _{B}( t) =\rho _{B}$. The state of
the system qubit is obtained by performing a partial trace over the bath
Hilbert space%
\begin{equation}
\rho _{S}( t) =\text{Tr}_{B}\{ U(t)\rho ( 0)U^{\dag
}(t)) \} ,  \label{System state}
\end{equation}%
where $U(t)=\exp [-it(H_{S}+H_{I}+H_{B})]$. The analytical solution of this
model was worked out in detail in Ref. \cite{Krovi:07}, and we present a
brief summary next.

\subsection{Exact Solution}

At any given time $t$, the state of the system qubit $\rho _{S}(
t) $ can be written in the Kraus representation as \cite{Kraus},%
\begin{equation}
\rho _{S}( t) =\mathcal{E}( \rho _{S}( 0) )
=\sum_{i,j}K_{ij}\rho _{S}( 0) K_{ij}^{\dag },
\label{Kraus representation}
\end{equation}%
where the Kraus operators satisfy the completeness relation $%
\sum_{i,j}K_{ij}^{\dag }K_{ij}=I_{S}$. After a transformation to the
interaction picture defined by $H_{S}+H_{B}$, these operators can be
expressed as%
\begin{equation}
K_{ij}=\sqrt{\lambda _{i}}\langle j| \exp (
-iH_{I}t) | i\rangle ,  \label{Kraus Operators}
\end{equation}%
where we have introduced the spectral decomposition $\rho
_{B}=\sum_{i}\lambda _{i}| i\rangle \langle i| 
$ of the initial bath state. For the Gibbs thermal state chosen here the
eigenbasis states $\{| i\rangle \}$ are $N$-fold tensor
products of the $\sigma ^{z}$ eigenstates, which gives%
\begin{equation}
\rho _{B}=\sum_{i}\frac{\exp ( -\beta E_{i}) }{Z}|
i\rangle \langle i| ,
\label{Initial state of bath in eigenbasis}
\end{equation}%
where $E_{i}=\frac{1}{2}\sum_{n=1}^{N}\Omega _{n}( -1) ^{i_{n}}$
is the energy of each bath eigenstate $| i\rangle$
($i=i_{1},i_{2},\ldots ,i_{N}$ is the binary expansion of the integer
$i$, where $i\in \lbrack 0,\ldots ,2^{N}-1]$) and $Z=\sum_{i}\exp ( -\beta
E_{i}) $ is the partition function. Therefore, the Kraus operators are%
\begin{equation}
K_{ij}=\sqrt{\lambda _{i}}\exp ( -it\alpha \widetilde{E}_{i}\sigma
^{z}) \delta _{ij},  \label{Ising Bath Kraus Operators}
\end{equation}%
with $\lambda _{i}=\exp ( -\beta E_{i}) /Z$, and $\widetilde{E}%
_{i}$ given by%
\begin{eqnarray}
\widetilde{E}_{i} &=&\sum_{n=1}^{N}g_{n}( -1) ^{i_{n}}-\text{Tr}%
\{ \sum_{n}g_{n}\sigma _{n}^{z}\rho _{B}\} ,  \notag \\
&=&\sum_{n=1}^{N}g_{n}[ ( -1) ^{i_{n}}-\beta _{n}] ,
\label{E Tilda}
\end{eqnarray}%
where $\beta _{n}=\tanh ( -\frac{1}{2}\beta \Omega _{n}) $. The
CPTP map $\rho _{S}( 0) \overset{\mathcal{E}}{\rightarrow }\rho
_{S}( t) $ with the Kraus operators given by Eq.~(\ref{Ising Bath
Kraus Operators}), represents a quantum dephasing channel, since the Kraus
operators are diagonal in the reference basis $\{|0\rangle ,|1\rangle \}\in 
\mathcal{H}_{S}$ (eigenstates of $\sigma ^{z}$ with eigenvalues $\pm 1$) of
the system. Moreover, as it corresponds to a non-Markovian model \cite%
{Krovi:07}, $\mathcal{E}$ represents a quantum dephasing channel with
memory. We now determine the information transmission capacities of this
channel.

\section{Capacities of Quantum Dephasing Channel}

\label{sec:comm}

\subsection{Classical Capacity}

Dephasing channels have the characteristic property of transmitting states
of a preferential orthonormal basis without introducing any error \cite%
{Nielsen:book}. These basis states can be used to encode classical
information, which makes these channels noiseless for the transmission of
classical information \cite{Arrigo:07}. Superpositions of the basis states
will decohere, however, therefore dephasing channels are noisy for quantum
information. For the dephasing channel ${\cal E}$ under consideration the
preferential orthonormal basis is$\{|0\rangle ,|1\rangle \}^{\otimes M}\in
{\cal H}_{S}^{\otimes M}$, for $M$ parallel uses of the channel, i.e., $M$
classical bits can be transmitted noiselessly over $M$ copies of the channel.

\subsection{Quantum Capacity}

Consider the communication system shown in Fig.~\ref{Communication protocol}. Quantum information is encoded  into the system spin
via a unitary  transformation. The system spin is then transmitted to the receiver, over
the spin-star channel. In general, one must perform the
maximization of the coherent information $I_{c}$ over
the $n$-fold
tensor product Hilbert space ${\cal H}_{S}^{\otimes n}$. However, Devetak and Shor
recently established dephasing channels as degradable channels
\cite{DevetakShor:05}.
Therefore the single channel-use formula $Q=Q_{1}$\
applies, and the maximization as in Eq.~(\ref{Quantum Capacity}) over the
larger Hilbert space is avoided. Moreover, Arrigo
{\it et al.} \cite{Arrigo:07}
showed that for dephasing channels the
coherent information $I_{c}$\ is
maximized by separable input states diagonalized in the reference
basis. Therefore, we set the initial state of the system spin as 
\begin{equation}
\rho _{S}( 0) =\frac{1}{2}( | 0\rangle
\langle 0| +| 1\rangle \langle
1| ) =\frac{I}{2}.  \label{Input state}
\end{equation}%
Initially, the system spin $\rho _{S}( 0) $\ is coupled to a
reference system $R$ such that the total system $SR$\ is pure. The
reference system does not undergo any dynamical evolution; it is
introduced as a mathematical device to purify the initial state of the
system spin. The joint initial state of the total system $SR$ is given
by the maximally entangled state
\begin{equation}
| \Phi \rangle =\frac{1}{\sqrt{2}}( |
00\rangle +| 11\rangle ) .  \label{Total System}
\end{equation}%
Dephasing channels are unital channels, i.e., $\mathcal{E}( I)
=I$, therefore the state of system spin is unaltered after interacting with
the Ising bath%
\begin{equation}
\rho _{S}( t) =\rho _{S}( 0) =\frac{I}{2}.
\label{Output state}
\end{equation}%
However, the total system $SR$ decoheres as a result of the interaction and
is mapped to a mixed state, whose diagonal elements (\textquotedblleft
populations\textquotedblright )\ are unaffected, but whose off-diagonal
elements (\textquotedblleft coherences\textquotedblright )\ are:%
\begin{eqnarray}
\rho _{SR}( t) &=&( \mathcal{E}\otimes \mathcal{I})
( | \Phi \rangle \langle \Phi | ) 
\notag \\
&=&\sum_{i,j}(K_{ij}\otimes I)( | \Phi \rangle
\langle \Phi | ) (K_{ij}^{\dag }\otimes I),  \notag \\
&=&\frac{1}{2}( | 00\rangle \langle 00|
+| 11\rangle \langle 11| )  \notag \\
&&+\frac{1}{2}\sum_{i}\lambda _{i}( e^{-2i\alpha t\widetilde{E}%
_{i}}| 00\rangle \langle 11|   \notag \\
&& +e^{+2i\alpha t\widetilde{E}_{i}}| 11\rangle
\langle 00| ) .  \label{Total system after interaction}
\end{eqnarray}%
The quantum capacity $Q$ of the dephasing channel
is now obtained by using Eq.~(\ref{Quantum Capacity}), making use of the
single channel-use formula $Q=Q_{1}$ and the fact that the coherent
information is maximized by our chosen initial state $\rho _{S}(
0)$:
\begin{eqnarray}
Q &=&Q_{1}=\max_{\rho _{S}\in \mathcal{H}_{S}}S[ \mathcal{E}( \rho
_{S}) ] -S[ ( \mathcal{E}\otimes \mathcal{I})
(|\Phi \rangle \langle \Phi |)]  \notag \\
&=&S[ \mathcal{E}( I/2) ]  \notag \\
&&-S[ ( \mathcal{E}\otimes \mathcal{I}) (\frac{1}{\sqrt{2}}%
( | 00\rangle +| 11\rangle ) \frac{1%
}{\sqrt{2}}( \langle 00|+\langle 11|) )]  \notag \\
&=&S[ I/2] -S[ \rho _{SR}( t) ] .
\end{eqnarray}%
This yields:
\begin{equation}
Q(t)=1+\sum_{k=1}^{4}{\large \chi }_{k}\log _{2}{\large \chi }_{k},  \label{Q}
\end{equation}%
where ${\large \chi }_{1}={\large \chi }_{2}=0$ and%
\begin{equation*}
{\large \chi }_{3}=\frac{1}{2}[ 1+\frac{1}{Z}|\Pi _{N}|] ,\quad 
{\large \chi }_{4}=\frac{1}{2}[ 1-\frac{1}{Z}|\Pi _{N}|] ,
\end{equation*}%
are the eigenvalues of the state $\rho _{SR}( t) $, and where 
\begin{equation}
\Pi _{N}(t)=\sum_{i=0}^{2^{N}-1}e^{-\sum_{n=1}^{N}( \frac{1}{2}\beta
\Omega _{n}+2i\alpha tg_{n}) ( -1) ^{i_{n}}}.  \label{Pi}
\end{equation}%
Next we calculate the entanglement-assisted capacities of the dephasing
channel.

\subsection{Entanglement-Assisted Capacities}

The communication protocol of entanglement-assisted capacities can
also be described using Fig.~\ref{Communication protocol}. Prior to the
communication the sender and receiver share a maximally entangled state
given by Eq.~(\ref{Total System}). The first qubit of the entangled pair
belongs to the sender: $\rho _{S}(0)={\rm Tr}_{R}(|\Phi \rangle \langle \Phi |) =I/2$, and
interacts with the bath. Unlike the quantum capacity protocol, the second
qubit is not a mathematical device and corresponds to the qubit in
possession of the receiver prior to the communication. Therefore, it is
again considered to have been transmitted over the identity channel.

Now note that in our case, since $S( \rho _{S}) =1$ and $Q=Q_{1}$, it
follows from Eqs.~(\ref{Quantum Capacity}) and (\ref{Entanglement-assisted classical capacity})
that the quantum capacity is related to the
entanglement-assisted classical capacity via the simple formula 
\begin{equation}
C_{E}=1+Q=2+\sum_{i=1}^{4}{\large \chi }_{i}\log _{2}{\large \chi }_{i},
\label{Exact Unlimited Ce}
\end{equation}
while the entanglement-assisted quantum capacity is
\begin{equation}
Q_{E}=\frac{C_{E}}{2}=1+\frac{1}{2}\sum_{i=1}^{4}{\large \chi }_{i}\log _{2}%
{\large \chi }_{i}.  \label{Exact Qe}
\end{equation}

Next, we are interested in the classical capacity assisted by limited
entanglement.\textit{\ }Consider the situation when instead of a maximally
entangled state, an ensemble of orthogonal states
\begin{eqnarray}
| \Phi_1\rangle  &=&\cos \theta |
00\rangle +\sin \theta | 11\rangle ,  \notag \\
| \Phi_2\rangle  &=&\sin \theta |
00\rangle -\cos \theta | 11\rangle ,  \notag \\
| \Phi_3\rangle  &=&\cos \theta |
01\rangle +\sin \theta | 10\rangle ,  \notag \\
| \Phi_4\rangle  &=&\sin \theta |
01\rangle -\cos \theta | 10\rangle ,
\label{Shared Ansatz}
\end{eqnarray}%
is shared prior to the communication, where $0\leq \theta \leq \frac{\pi }{4}
$. As above the first and second qubits belong to the sender and receiver,
respectively. We show in Appendix \ref{app} that the classical capacity
assisted by limited entanglement [Eq.~(\ref{Classical capacity with limited
entanglement})] is attained when all states $\{| \Phi_i\rangle
\}_{i=1}^{4}$ are equiprobable, and that this yields:%
\begin{eqnarray}
C_{E}^{\lim } &=&-[ \cos ^{2}\theta \log _{2}\cos ^{2}\theta +\sin
^{2}\theta \log _{2}\sin ^{2}\theta ]+1   \notag \\
&&+\sum_{i=1}^{4}\omega _{i}\log _{2}\omega _{i},  \label{Exact limited Ce}
\end{eqnarray}%
with $\omega _{1}=\omega _{2}=0$ and%
\begin{eqnarray}
\omega _{3} &=&\frac{1}{2}[ 1+\{ (\frac{2\cos\theta \sin\theta }{Z}|\Pi _{N}|)^{2}+\cos ^{2}2\theta \} ^{\frac{1}{2}}%
] ,  \notag \\
\omega _{4} &=&\frac{1}{2}[ 1-\{ (\frac{2\cos\theta \sin\theta }{Z}|\Pi _{N}|)^{2}+\cos ^{2}2\theta \} ^{\frac{1}{2}}%
] .  \label{Limited Entanglement Eigenvalues}
\end{eqnarray}%
For $\theta =0$ the states given by Eq.~(\ref{Shared Ansatz}) are product
states and we recover the classical capacity which is equal to one. The
capacity $C_{E}^{\lim }$ increases as we increase the value of $\theta $,
attaining its maximum for $\theta =\frac{\pi }{4}$ for which the states are
maximally entangled and Eq.~(\ref{Exact limited Ce}) reduces to Eq.~(\ref%
{Exact Unlimited Ce}).

\begin{figure}[tph]
\centering
\includegraphics[width=3.4in]{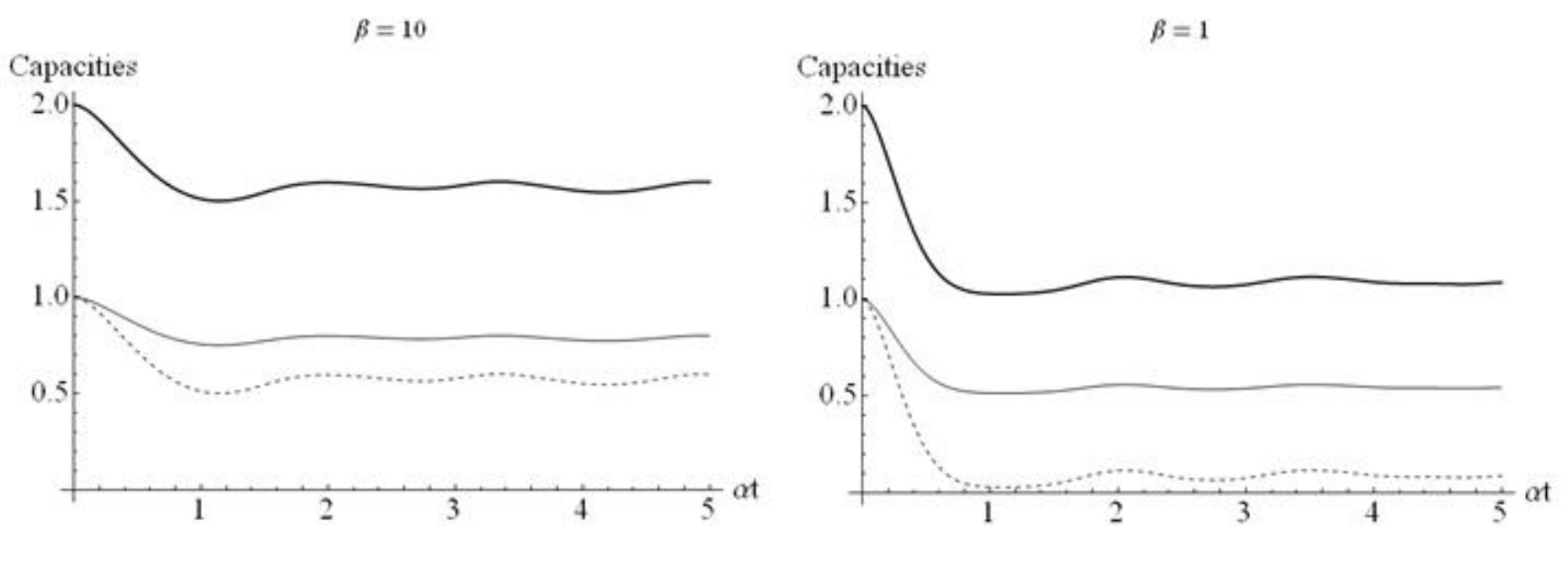}
\caption{Capacities $C_{E}=1+Q$ (solid, thick), $Q_{E}=C_{E}/2$ (solid,
thin) and $Q$ (dashed) of qubit coupled to an Ising spin bath with $N=4$,
for random values of $g_{n}$ and $\Omega _{n}$. Left: $\beta=10$,
right: $\beta=1$. See text for details.}
\label{Fig 2}
\end{figure}

\begin{figure}[tph]
\centering
\includegraphics[width=3.4in]{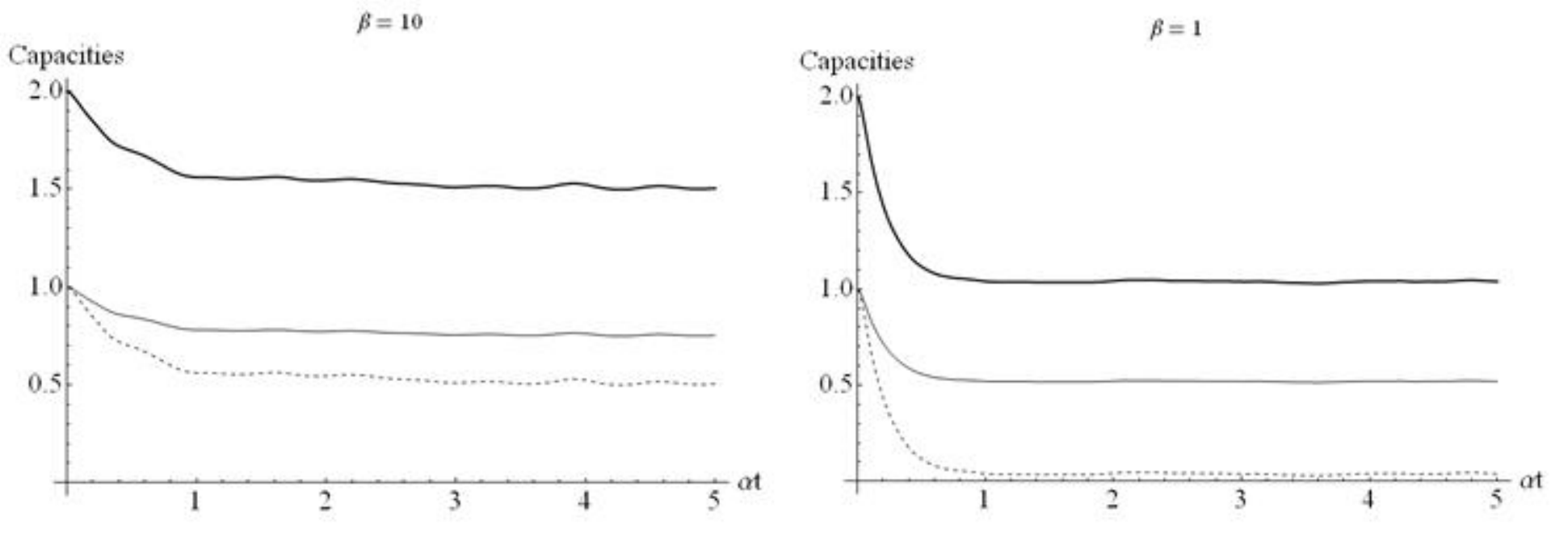}
\caption{Same as Fig.~\protect\ref{Fig 2}, with $N=100$.}
\label{Fig 3}
\end{figure}

Plots of capacities for random values of couplings $g_{n}$ and bath
frequencies $\Omega _{n}$, are given in Figs.~\ref{Fig 2} and \ref{Fig 3}. We
generate real, random values of $g_{n}$ and $\Omega _{n}$ uniformly
distributed in the interval $[ -1,1] $ and plot average
capacities for $50$ random ensembles. In Fig.~\ref{Fig 2}, we plot the
capacities of the system spin coupled to a bath with $N=4$ spins. We plot
the capacities at low and high temperatures in order to study the effect of
bath temperature. For low temperature ($\beta =10$), the bath is not too
noisy and the system spin retains its coherence well. The capacities do not
acquire their minimum values and partial recurrences occur, with an
amplitude that diminishes over time.
At high temperature ($\beta =1$) the capacities rapidly decrease to
their minimum values and the recurrences are of smaller amplitude. As the
system spin loses its coherence to the Ising bath, the entanglement shared
between the sender and receiver is destroyed and $C_{E}$ is reduced to its
minimum value of one. This corresponds to the qubit in possession of the
receiver prior to the communication. The quantum capacity $Q$, which is a
measure of the coherent information transmitted, is reduced to zero as the
system spin decoheres completely. As we increase the number of bath spins to 
$N=100$, we observe a similar dependence on bath temperature. The main
difference compared to the case of a small number of bath spins is the
drastically diminished amplitude of the recurrences. As
noted in Ref.~\cite{Krovi:07}, this behavior is due to the averaging of the
positive and negative oscillations which arise for different values of the
parameters $g_n$ and $\Omega_n$.

\subsection{Limiting Cases}

\subsubsection{Equal Couplings and Frequencies}

\begin{figure}[tph]
\centering
\includegraphics[width=3.4in]{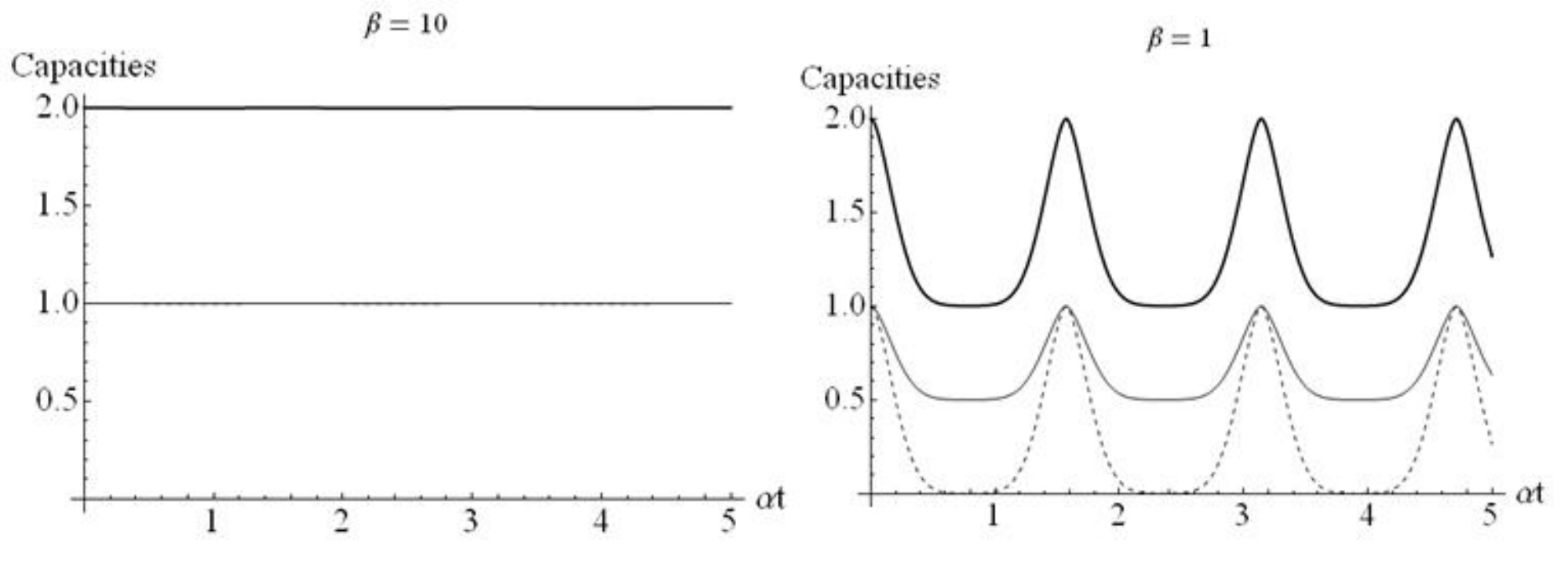}
\caption{Capacities $C_{E}$ (solid, thick), $Q_{E}$ (solid, thin) and $Q$
(dashed) of a qubit coupled to an Ising spin bath with $N=4$, for $g_{n}=1$ and $%
  \Omega _{n}=1$ $\forall n$. Left: $\beta=10$, right: $\beta=1$.}
\label{Fig 4}
\end{figure}

\begin{figure}[tph]
\centering
\includegraphics[width=3.4in]{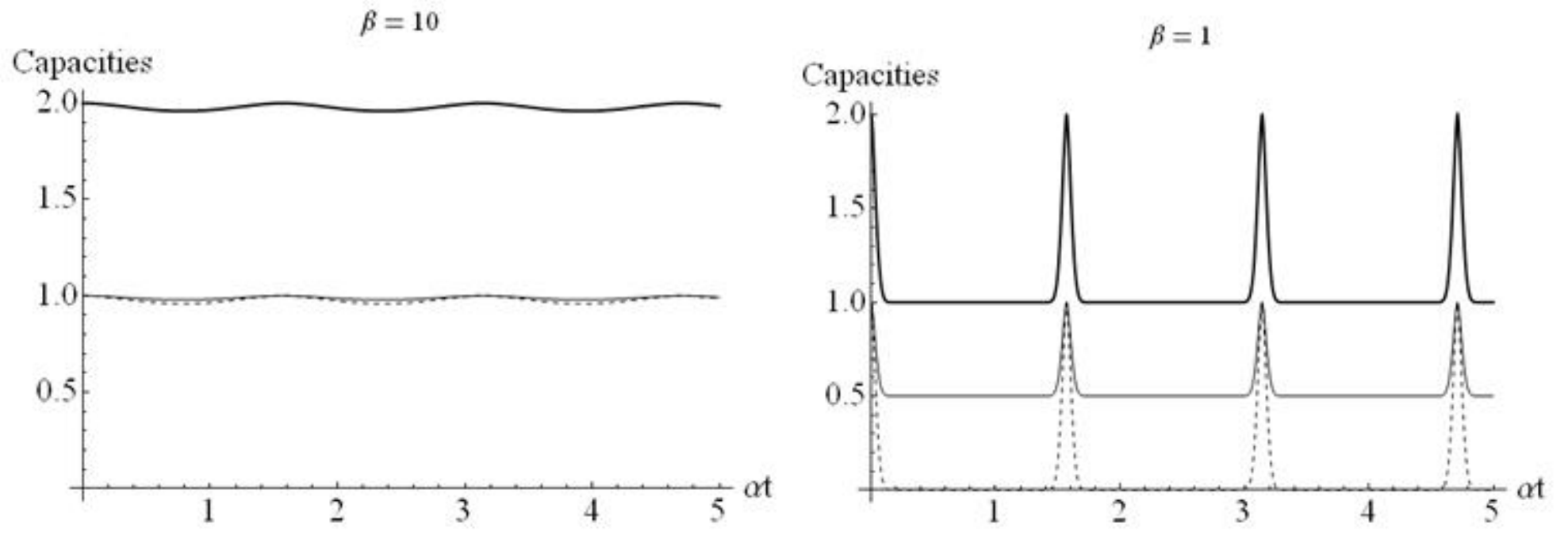}
\caption{Same as Fig.~\protect\ref{Fig 4}, with $N=100$.}
\label{Fig 5}
\end{figure}

If the bath spins have equal frequencies $\Omega _{n}\equiv \Omega $ $%
\forall n$, and couplings $g_{n}\equiv g$ $\forall n$ with the system spin
then Eq.~(\ref{Pi}) reduces to 
\begin{eqnarray}
\Pi _{N} &=&\sum_{i=0}^{2^{N}-1}e^{-f(t)\sum_{n=1}^{N}( -1)
^{i_{n}}}  \notag \\
&=&\sum_{k=0}^{N}\binom{N}{k}e^{(2k-N)f(t)}=(2\cosh [f(t)])^{N},
\label{Equal}
\end{eqnarray}%
where 
\begin{equation}
f(t)=\frac{1}{2}\beta \Omega +2i\alpha tg.
\end{equation}%
The second equality in Eq.~(\ref{Equal}) follows from the fact that the term 
$\sum_{n=1}^{N}( -1) ^{i_{n}}=N-2k$ for $i$ with Hamming weight $%
k $, of which there are $\binom{N}{k}$ cases for $i\in \lbrack 0,\ldots
,2^{N}-1]$. Therefore $|\Pi _{N}|=2^{N}\sqrt{(\cos ^{2}(2\alpha tg)\cosh
^{2}(\frac{\beta }{2}\Omega )+\sin ^{2}(2\alpha tg)\sinh ^{2}(\frac{\beta }{2%
}\Omega ))^{N}}$ is periodic with period $T_{p}=\pi /(2\alpha g)$, and the
same is true of all the capacities computed above. At these times the bath
spins destructively interfere and the dephasing channel becomes noiseless
for information transmission. In the high temperature limit $\beta \Omega
\rightarrow 0$, and $|\Pi _{N}|\rightarrow (2\cos (2\alpha tg))^{N}$, so
that the capacities exhibit full periodic recurrences independently of $N$,\
in contrast to the results for random couplings and frequencies. Clearly, as 
$N$ gets larger, these recurrences become sharper, until in the limit $%
N\rightarrow \infty $ they become isolated peaks, as shown in the
right-side panels of
Figs.~\ref{Fig 4} and \ref{Fig 5}. In the low temperature limit $\beta \Omega \rightarrow
\infty $ and $|\Pi _{N}|\rightarrow \exp (\frac{N}{2}\beta \Omega )$, but so
does the partition function $Z=\sum_{i=0}^{2^{N}-1}e^{-\frac{1}{2}\beta
\Omega \sum_{n=1}^{N}( -1) ^{i_{n}}}=(2\cosh [f(0)])^{N} \rightarrow \exp (\frac{N}{2}\beta \Omega )$, so
that ${\large \chi }_{3},{\large \chi }_{4}\rightarrow 1$, and all the
capacities are saturated at their maximum values. For small, but
finite temperatures, the capacities exhibit oscillations with an
amplitude that grows with $N$, as can be seen in the left-side panels of
Figs.~\ref{Fig 4} and \ref{Fig 5}. This is in contrast to the case of
random couplings seen in Figs.~\ref{Fig 2} and \ref{Fig 3}; there
destructive interference caused a cancellation of these oscillations,
while in the case of equal couplings the capacity oscillations
survive and grow with the number of bath spins, reflecting the increased information
transfer from the system to the bath as a function of bath size.

\subsubsection{Large $N$}

Without symmetries in the coupling constants or frequencies the capacities
rapidly decrease to their minimum values and we find no recurrences for $%
N\gg 1$, high temperature and uniformly distributed random values of $g_{n}$
and $\Omega _{n}$. However, partial recurrences occur in this situation for
small temperature.

\subsubsection{Short Times}

The capacities are flat initially and do not decay exponentially in the
limit of short times $\alpha t\ll 1$, provided that the temperature and
number of bath spins $N$ is not too large. This is somewhat similar to the
Zeno behavior pointed out in Ref.~\cite{Krovi:07}.

\section{Summary and Conclusions}

\label{sec:conc}

We have studied an exactly solvable spin-star system for transmission of
classical and quantum information. The information is encoded into a system
spin which interacts with a spin-bath via arbitrary Ising couplings. We
considered the \textquotedblleft parallel uses\textquotedblright\ setting of
a memoryless quantum channel, where multiple copies of the system spin are
transmitted simultaneously via the same number of copies of the spin bath.
As our model is described by the dephasing channel, classical information
can be transmitted noiselessly, while the quantum capacity can be determined
via the \textquotedblleft single-letter\textquotedblright\ formula $Q=Q_{1}$%
, i.e., it suffices to consider a single copy of the spin-star system. We
analytically determined the quantum capacities of this communication system,
which exhibit a strong dependence on the couplings of the bath spins with
the system, and on the bath temperature. The Ising spin bath becomes noisier
as the temperature is increased, and the capacities rapidly deteriorate. For random
couplings and frequencies, recurrences are of
small amplitude and die out rapidly. However, for
equal couplings and frequencies full periodic recurrences occur
independently of the number of bath spins. These recurrences are a signature
of the non-Markovian nature of the spin-bath. At low temperature the quantum
capacities remain high when the number of bath spins is not too large.

\begin{acknowledgments}
N.A. was supported by Higher Education Commission Pakistan under grant no.
063-111368-Ps3-001 and IRSIP-4-Ps-08. D.A.L. was supported by the National
Science Foundation under grants no. PHY-803304 and CHE-924318.
\end{acknowledgments}

\appendix{}

\section{Derivation of the result for the classical capacity assisted by limited entanglement}
\label{app}

We prove Eq.~(\ref{Exact limited Ce}). Without loss of generality the ensemble of
orthogonal states given by Eq.~(\ref{Shared Ansatz}) can be assumed to
appear with probabilities parametrized as 
\begin{eqnarray}
p_{1}( x_{1},x_{2})  &=&\cos ^{2}x_{1}\cos ^{2}x_{2},  \notag \\
p_{2}( x_{1},x_{2})  &=&\sin ^{2}x_{1}\cos ^{2}x_{2},  \notag \\
p_{3}( x_{1},x_{2})  &=&\cos ^{2}x_{1}\sin ^{2}x_{2},  \notag \\
p_{4}( x_{1},x_{2})  &=&\sin ^{2}x_{1}\sin ^{2}x_{2},
\label{Probability Distribution}
\end{eqnarray}%
where $0\leq \theta ,x_{1},x_{2}\leq \frac{\pi }{4}$. We will show by
explicit calculation that for our pure dephasing model only the second term in Eq.~(\ref%
{Classical capacity with limited entanglement}) for the classical capacity
assisted by limited entanglement depends on the parameters $x_{1},x_{2}$.
Hence the maximization can be carried out, for fixed $\theta $, by
maximizing only this second term.

The states input to the quantum dephasing channel obtained from Eq.~(\ref%
{Shared Ansatz}) are
\begin{eqnarray}
\rho _{S,1} &=&\text{Tr}_{R}( |\Phi_1\rangle \langle \Phi_1 |) =\cos ^{2}\theta | 0\rangle
\langle 0| +\sin ^{2}\theta | 1\rangle
\langle 1| ,  \notag \\
\rho _{S,2} &=&\text{Tr}_{R}( |\Phi_2\rangle \langle \Phi_2 |) =\sin ^{2}\theta | 0\rangle
\langle 0| +\cos ^{2}\theta | 1\rangle
\langle 1| ,  \notag \\
\rho _{S,3} &=&\text{Tr}_{R}( |\Phi_3\rangle \langle \Phi_3 |) =\cos ^{2}\theta | 0\rangle
\langle 0| +\sin ^{2}\theta | 1\rangle
\langle 1| ,  \notag \\
\rho _{S,4} &=&\text{Tr}_{R}( |\Phi_4\rangle \langle \Phi_4|) =\sin ^{2}\theta | 0\rangle
\langle 0| +\cos ^{2}\theta | 1\rangle
\langle 1| ,\notag \\
\end{eqnarray}%
therefore, for all $\rho _{S,i}$%
\begin{equation}
S( \rho _{S,i}) =-[ \cos ^{2}\theta \log _{2}\cos ^{2}\theta
+\sin ^{2}\theta \log _{2}\sin ^{2}\theta ] .  \label{Input entropy}
\end{equation}%
This results in the following expression for the first term in
Eq.~(\ref{Classical capacity with limited entanglement}): 
\begin{align}
&\sum_{i}p_{i}( x_{1},x_{2}) S( \rho _{S,i})
\notag \\
&=-( \cos ^{2}x_{1}\cos ^{2}x_{2}+\sin ^{2}x_{1}\cos ^{2}x_{2} 
\notag \\
& \quad +\cos ^{2}x_{1}\sin ^{2}x_{2}+\sin ^{2}x_{1}\sin
^{2}x_{2}) \notag \\
&\quad \times [ \cos ^{2}\theta \log _{2}\cos ^{2}\theta +\sin ^{2}\theta
\log _{2}\sin ^{2}\theta ] ,  \notag \\
&=-[ \cos ^{2}\theta \log _{2}\cos ^{2}\theta +\sin ^{2}\theta \log
_{2}\sin ^{2}\theta ] ,  \label{first term}
\end{align}
where we have used the normalization $\sum_{i=1}^{4}p_{i}(
x_{1},x_{2}) =1,\forall x_{1},x_{2}$. This yields the
first term in Eq.~(\ref{Exact limited Ce}).

Next we calculate the second
term in Eq.~(\ref{Classical capacity with limited entanglement}). The output
state is 
\begin{align}
&\mathcal{E}( \sum_{i}p_{i}( x_{1},x_{2}) \rho
  _{S,i}) \notag \\
&=\sum_{i,j}K_{ij}( \sum_{i}p_{i}( x_{1},x_{2}) \rho
_{S,i}) K_{ij}^{\dag },  \label{output state}
\end{align}
where%
\begin{align}
&\sum_{i}p_{i}( x_{1},x_{2}) \rho _{S,i} \notag \\
&=[ \cos ^{2}\theta
( \cos ^{2}x_{1}\cos ^{2}x_{2}+\cos ^{2}x_{1}\sin ^{2}x_{2})
   \notag \\
& \quad +\sin ^{2}\theta ( \sin ^{2}x_{1}\sin ^{2}x_{2}+\sin
^{2}x_{1}\cos ^{2}x_{2}) ] | 0\rangle
\langle 0|   \notag \\
& \quad +[ \cos ^{2}\theta ( \sin ^{2}x_{1}\sin ^{2}x_{2}+\sin
^{2}x_{1}\cos ^{2}x_{2})    \notag \\
& \quad  +\sin ^{2}\theta ( \cos ^{2}x_{1}\cos ^{2}x_{2}+\cos
^{2}x_{1}\sin ^{2}x_{2}) ] | 1\rangle
\langle 1|   \notag \\
&=( \cos ^{2}\theta \cos ^{2}x_{1}+\sin ^{2}\theta \sin
^{2}x_{1}) | 0\rangle \langle 0|   \notag
\\
& \quad +( \cos ^{2}\theta \sin ^{2}x_{1}+\sin ^{2}\theta \cos
^{2}x_{1}) | 1\rangle \langle 1| .
\label{input ensemble}
\end{align}
Since this state is diagonal (``classical'') it is invariant under the dephasing channel with Kraus operators given by
Eq.~(\ref{Ising Bath Kraus Operators}). Therefore the eigenvalues of the
output state (\ref{output state}) are
\begin{eqnarray}
\upsilon _{1} &=&\cos ^{2}\theta \cos ^{2}x_{1}+\sin ^{2}\theta \sin
^{2}x_{1},  \notag \\
\upsilon _{2} &=&\cos ^{2}\theta \sin ^{2}x_{1}+\sin ^{2}\theta \cos
^{2}x_{1},  \label{Output eigenvalues}
\end{eqnarray}%
and%
\begin{equation}
S[ \mathcal{E}( \sum_{i}p_{i}( x_{1},x_{2}) \rho
_{S,i}) ] =-\sum_{i=1}^{2}\upsilon _{i}\log _{2}\upsilon _{i}.
\label{output entropy}
\end{equation}

Finally, we calculate the third term in Eq.~(\ref{Classical capacity with
limited entanglement}):%
\begin{align}
&( \mathcal{E}\otimes I) (|\Phi_i\rangle \langle \Phi_i|)\notag \\
&=\sum_{i,j}(K_{ij}\otimes I)( |\Phi_i\rangle \langle \Phi_i|) (K_{ij}^{\dag }\otimes I),
\label{purification output}
\end{align}
which for all $|\Phi_i\rangle $ has eigenvalues $\omega
_{1}=\omega _{2}=0$ and $\omega _{3},\omega _{4}$ are given in Eq.~(\ref%
{Limited Entanglement Eigenvalues}). The third term in Eq.~(\ref{Classical
capacity with limited entanglement}) is thus%
\begin{align}
&\sum_{i}p_{i}( x_{1},x_{2}) S[ ( \mathcal{E}\otimes
I) (|\Phi_i\rangle \langle \Phi_i|)] \notag \\
&=-\sum_{i=1}^{4}\omega _{i}\log _{2}\omega _{i},  \label{third term}
\end{align}
where, as for Eq.~(\ref{first term}), we have used the normalization $\sum_{i=1}^{4}p_{i}=1$.
This yields the third term in Eq.~(\ref{Exact limited Ce}).

Thus, indeed only the second term in Eq.~(\ref{Classical capacity with
limited entanglement}) depends on $x_{1},x_{2}$, and for a given value of $%
\theta $, the classical capacity assisted by limited entanglement is
maximized by maximizing Eq.~(\ref{output entropy}). The maximum is attained
when the output state (\ref{output state}) is fully mixed, i.e., when its
eigenvalues $\upsilon _{1}=\upsilon _{2}=1/2$. This occurs when $x_{1}=x_{2}=%
\frac{\pi }{4}$, i.e., when we have an equiprobable ensemble of the
states. This gives rise to the $1$ in Eq.~(\ref{Exact limited Ce}).


\end{document}